\documentclass[showpacs,preprintnumbers,amsmath,amssymb,preprint]{revtex4}
\usepackage{graphics}
\begin{document}

\title{Microcanonical studies on isoscaling}
\author{Ad. R. Raduta}
\affiliation{National Institute of Nuclear Physics and Engineering,\\
Bucharest, POB MG6, Romania}

\begin{abstract}
The exponential scaling of isotopic yields is investigated 
for sources of different sizes over a broad range 
of excitation energies and freeze-out volumes, in both primary and asymptotic 
stages of the decay in the framework of a microcanonical 
multifragmentation model. 
It was found that the scaling parameters have a strong dependence on the 
considered pair of equilibrated sources and excitation energy and are
affected by the secondary particle emission of the break-up fragments. No
significant
influence of the freeze-out volume on the considered isotopic ratios has been
observed. Deviations of microcanonical results from grandcanonical expectations are discussed.
\end{abstract}


\maketitle

\section{Introduction}

Many nuclear physics experiments realized in the last years revealed an 
interesting scaling property of the isotopic yield ratios obtained from the 
disintegration of equilibrated sources of similar sizes, 
with close values of excitation energy per
nucleon and temperature but different isospin values as a function of the
isotopic composition of the emitted cluster \cite{Xu,Tsang_RC2001}.  
This property has been called
isoscaling and can be expressed mathematically by the formula,

\begin{equation}
R_{21}(Z,N)=Y_2(Z,N)/Y_1(Z,N)= C \exp(\alpha N+\beta Z),
\label{eq.isoscaling}
\end{equation}
where $Y_i(Z,N)$ denotes the yield of the isotope $(Z,N)$ obtained from
the decay of the excited nuclear system $"i"$ and $C$, $\alpha$ and $\beta$
are scaling parameters.

If one classifies the nuclear reactions which manifest isoscaling function
of the projectile energy, one can
say that isoscaling has been observed in reactions induced by projectiles
whose energies range from a few MeV/nucleon to  several GeV/nucleon. In terms of
reaction mechanisms, isoscaling has been obtained in evaporation reactions
(e.g. $^4$He + $^{116}$Sn, $^{124}$Sn at 50 MeV/nucleon bombarding energy
\cite{evap}),
deep inelastic reactions (e.g. $^{16}$O + $^{232}$Th, $^{16}$O + $^{197}$Au at 8.6
MeV/nucleon \cite{deep-inelastic}, $^{86}$Kr + $^{112,124}$Sn
and $^{86}$Kr + $^{58,64}$Ni at 25 MeV/nucleon \cite{souliotis}) 
and, recently,
in fission reactions (e.g. n + $^{233,238}$U at 14 MeV/nucleon
\cite{fission}). The biggest amount of experimental isoscaling data comes
from multifragmentation reactions 
\cite{Xu,Tsang_RC2001,Tsang_prl2001,Botvina,Chimera,shetty}
and the great interest in these studies
was motivated by the possibility to infer information on the nuclear equation
of state, and in particular on the asymmetry term,
from the isoscaling parameters \cite{Botvina,Tsang_stat_mod}.

It is interesting to specify at this point that, relatively easy to understand
for nuclear reactions involving equilibrated compound nuclear systems 
\cite{Tsang_prl2001},
isoscaling is compatible also with dynamical fragment formation as was
recently demonstrated in Ref. \cite{ono,Liu}.

The aim of the present paper is to investigate isoscaling in multifragmentation 
reactions and more precisely the results produced within a microcanonical
framework. 
In order to learn how sensitive the isoscaling parameters are with respect
to the observables characterizing the source's state, we considered
sources whose nucleon number varies from 40 to 200, with excitation
energies between 2 and 15 MeV/nucleon, freeze-out volumes ranging
from $3V_0$ to $10V_0$ in both primary and asymptotic stages of the decay.  
The relevance of such study consists in the fact
that the most suitable statistical ensemble to describe the decay of an isolated 
excited nuclear system
counting at most several hundreds of nucleons is the microcanonical ensemble.
The importance of using models which can mimic as well as possible
the real physical case
is demonstrated by the fact that the best
description of the experimental data was obtained by the
predictions of microcanonical multifragmentation models 
\cite{SMM,MMMC,Noi_Indra}.

Section \ref{sec.iso} makes a review of the grandcanonical argumentation 
of isoscaling in nuclear multifragmentation.
Section \ref{sec.micro} presents microcanonical results on isoscaling
obtained for different mass systems with excitation
energies ranging from 2 to 15 MeV/nucleon at different freeze-out volumes
in both primary and asymptotic stages of the decay. The sensitivity of the isoscaling 
parameters to observables characterizing the state of the source is discussed in detail.
Deviations of microcanonical results from grandcanonical expectations are commented. 
The conclusions are drawn in Section \ref{sec.concl}.

\section{Isoscaling in nuclear multifragmentation}
\label{sec.iso}

Isoscaling in multifragmentation was for the first time evidenced for the 
light emitted fragments ($A\leq 18$) obtained in the reactions 
$^{112}$Sn+$^{112}$Sn and $^{124}$Sn+$^{124}$Sn at 50 MeV/nucleon 
bombarding energy \cite{Xu} and confirmed by many other 
reactions induced by relativistic light particles 
(e.g. p, d, $\alpha$+$^{112}$Sn and $^{124}$Sn at incident energies
between 660 MeV and 15.3 GeV \cite {Botvina})
or heavy ions ($^{124}$Sn+$^{64}$Ni and $^{112}$Sn+$^{58}$Ni 
at 35 MeV/nucleon \cite{Chimera}).

The theoretical interpretation of isoscaling in the multifragmentation energy
domain was given using a grandcanonical ensemble \cite{Tsang_prl2001}.
It was demonstrated that in 
the grandcanonical limit the yield of a primary
fragment $(Z,N)$ obtained from the decay of a thermally equilibrated system 
characterized by the temperature $T$ can be written as \cite{albergo}:

\begin{equation}
Y(Z,N) \propto \exp(B(Z,N)/T)\exp(N \mu_n/T+Z\mu_p/T),
\label{eq.yield}
\end{equation}
where $B(Z,N)$ is the binding energy of the emitted fragment $(Z,N)$ and $\mu_n$ 
and $\mu_p$ are the neutron and proton chemical potentials.
Taking into account that the two equilibrated systems have the same temperature, 
excitation energy per nucleon and similar sizes the
proportionality factors from Eq. (\ref{eq.yield}) have been considered practically
identical in the two cases.
Thus, one obtains for the ratio of the yields of any isotope resulting 
from the decay of the two sources the expression of Eq. (\ref{eq.isoscaling}) 
where,

\begin{equation}
\alpha=\Delta \mu_n/T,
\label{eq.mu_n}
\end{equation}
and
\begin{equation}
\beta=\Delta \mu_p/T,
\label{eq.mu_p}
\end{equation}
where $\Delta \mu_{n(p)}=\mu_{2n(p)}-\mu_{1n(p)}$.

Despite the fact that the suitability of a grandcanonical approximation
is hard to be justified for systems having at most few hundreds particles,
the success of the above formulas was due to their simplicity
and to the fact that no contradiction came from experimental data 
or various statistical multifragmentation models (e.g. microcanonical \cite{SMM} and 
canonical \cite{SMM_gc} versions of SMM, Expanding Emitting Source \cite{EES}) 
currently used in data analyzes \cite{Tsang_stat_mod}.

Concerning the slope parameters $\alpha$ and $\beta$, several values have been
reported in literature. This fact is expected because is known that 
particle formation probabilities have explicit dependence of
all observables which characterize an excited source (mass $A$, charge $Z$, 
excitation energy $E_{ex}$, freeze-out volume $V$). It is also known that
the sequential particle emission from the excited primary fragments modifies
the isotopic yields. This evaporation process may also affect the values of
$\alpha$ and $\beta$.

\section{Isoscaling from a microcanonical perspective}
\label{sec.micro}

Despite the fact that, contrary to the grandcanonical case, no microcanonical model
provides analytical expressions for cluster yields, the occurrence of isoscaling
in a microcanonical model can be understood starting from the shape of the isotopic 
yields corresponding to a given element \cite{Botvina}.
Thus, it has been observed that charge distributions of fragments with fixed 
values of the neutron number ($Y(Z,N)|_N$) and $N$ distributions of 
fragments with fixed 
values of the proton number ($Y(Z,N)|_Z$) can be approximated 
by Gaussian functions.
The occurrence of such Gaussian-like distributions within a microcanonical model
can be regarded as a natural consequence of complete equilibration.
Using the generic notation $Y(x,y)|_y$ for the distribution  of the observable 
$x$ when
$y$ is kept fixed (where $x=N,Z$ and $y=Z,N$), one can write,

\begin{equation}
Y(x,y)|_y=C \exp(-(x-x_{\sf{med}}(y))^2/2\sigma(y)^2).
\label{yields}
\end{equation}

The mean value $x_{\sf{med}}(y)$ gives the isotope produced with the largest
probability, while the variance $\sigma(y)$ is a measure of how uniform is the
population of fragments having different isospin values.
The statistical meaning of these quantities involves an obvious dependence on
observables determining the ensemble's state. Since we deal with a microcanonical approximation, 
these observables are the
mass, volume and excitation energy of the multifragmenting source.

Using Eq. \ref{yields}, the ratio of isotopic yields writes,

\begin{eqnarray}
\label{eq.geom}
\nonumber
R_{21}(x,y={\sf const.})=\frac{Y_2(x,y)|_y}{Y_1(x,y)|_y}\\
=\exp\left[
-{\frac{x^2}{2} \left(\frac{1}{\sigma_2(y)^2-\sigma_1(y)^2}\right)}
+{x \left(\frac{x_{{\sf med}2}(y)}{\sigma_2(y)^2}-\frac{x_{{\sf med}1}(y)}{\sigma_1(y)^2}\right)}
-{\left(\frac{x_{{\sf med}2}(y)^2}{2\sigma_2(y)^2}-\frac{x_{{\sf med}1}(y)^2}{2\sigma_1(y)^2}\right)}
\right].
\end{eqnarray}

Taking into account that the two considered sources have similar sizes, temperatures
and freeze-out volumes, the isotopic distributions corresponding to any
cluster are expected to have equal variances
($\sigma_1(y)=\sigma_2(y)=\sigma(y)$).
Thus, in Eq. (\ref{eq.geom}) the quadratic term in $x$ vanishes 
leading to a linear dependence of the yield ratio on the observable
$x$ of the emitted cluster. 

From Eq. (\ref{eq.geom}) one can express the $\alpha$
and $\beta$ slope parameters in terms of mean values and variances of isotopic 
distributions,
\begin{equation}
\alpha(Z)=\frac1{\sigma(Z)^2}\left(N_{{\sf med}2}(Z)-N_{{\sf med}1}(Z)\right),
\label{eq.alpha.g}
\end{equation}
\begin{equation}
\beta(N)=\frac1{\sigma(N)^2} \left(Z_{{\sf med}2}(N)-Z_{{\sf med}1}(N)\right).
\label{eq.beta.g}
\end{equation}

From Eqs. (\ref{eq.alpha.g}) and (\ref{eq.beta.g}) results that $\alpha$ and $\beta$
may depend on the size of the emitted cluster.
A necessary condition to obtain
slope parameters independent of the size of the emitted cluster as in Eq. \ref{eq.isoscaling}, 
is to have a constant value for the ratio
of the shift between the isotopic distributions produced by the two
sources and the square of the distributions' variance. As we shall see in the following
sections this hold with good approximation for the light clusters emitted
by relatively large equilibrated sources but does not hold for heavy fragments 
emitted by heavy sources and clusters emitted by light multifragmenting sources.

In the following we shall investigate isoscaling over a
large range of masses, excitation energies and freeze-out volumes of the 
multifragmenting sources in order to determine the dependence of 
the slope
parameters $\alpha$ and $\beta$ on the source parameters, the influence of
the secondary decays on $\alpha$ and $\beta$ and, in the case of small sources, 
the dependence of $\alpha$ and $\beta$ on the size of the emitted cluster.
To serve this goal we use the microcanonical multifragmentation model presented in detail
in Refs.\cite{Noi,Noi_Indra}.

\subsection{Model overview} 

The microcanonical multifragmentation model \cite{Noi,Noi_Indra} has two
distinct stages: the break-up stage and the sequential particle evaporation 
stage.

The most important part of the model is the so called break-up stage which aims to 
describe the explosion of the equilibrated nuclear source. In order to mimic as well as possible the 
physical situation of a small isolated system with fixed excitation energy, the model
tries to obey sharply to microcanonical constrains. Thus, the fixed observables characterizing the state
of the equilibrated source are mass, charge,
total energy $E$, total momentum ${\bf P}$ (=0 in the center of mass (c.m.) 
frame) and total angular momentum ${\bf L}$ (=0 for non-rotating systems).
The freeze-out volume $V$ is determined by the spherical container in which fragments are generated. 

The standard version of the model allows treatment of freeze-out volume according to two
different scenarios \cite{Noi_Indra}: hard spherical nonoverlapping fragments placed 
in a spherical container and spherical container with free volume parameterization 
\cite{vol-randrup}.
For a faster simulation we considered the second case. The mathematical expression of the 
free volume is:
\begin{equation}
V_{free}=\prod_{i=1}^{N_{fr}} (V-i \frac{V_0}{N_{fr}}),
\end{equation}
where $N_{fr}$ is the number of fragments in a given event and $V_0$ is the 
volume of the nuclear system at normal density.

The price to pay for a model aiming to respect all microcanonical constrains is the impossibility to obtain
formulas analytically tractable. The key quantity of the model is the statistical weight of a 
configuration $C$,

\begin{eqnarray}
W_C \propto \frac1{N_C!}\prod_{n=1}^{N_C}\left(\Omega \frac{\rho_n(\epsilon_n)}{h^3}(mA_n)^{3/2}\right)
\times \frac{2\pi}{\Gamma(3/2(N_C-2))}\frac{1}{\sqrt{({\rm det} I})}
\frac{(2 \pi K)^{3/2N_C-4}}{(mA)^{3/2}},
\end{eqnarray}
where $N_C$ is the number of fragments corresponding to configuration $C$,
$\Omega$ is the accesible volume,
$I$ is inertial tensor of the system and $K$ is the available kinetic energy.
The index $n$ denotes the fragments in each configuration. 

The average value of any observable $Q$ is calculated via a Metropolis type simulation 
according to the expression,

 \begin{equation}
<Q>=\frac{\sum_C Q_C W_C}{\sum_C W_C}.
 \end{equation}
 
Because the observables studied in the present paper are functions of isotopic
yields they are strongly dependent on all specific details of the model
(binding energy parameterization, allowed isospin asymmetry of primary fragments, 
employed level density formula, etc.). 
The break-up fragments allowed to be formed are all isotopes included in the
mass table of Ref. \cite{mass}.

The fragments' binding energy was calculated according to the liquid-drop
parameterization:

\begin{eqnarray}
\nonumber
B(A,Z)&=&15.4941 (1-1.7826 I^2) A\\
\nonumber
&-&17.9439 (1-1.7826 I^2) A^{2/3}\\
&-&0.7053 Z^2 A^{-1/3}+1.1530 Z^2/A,
\end{eqnarray}
where $I=(A-2Z)/A$.

Fragments with mass smaller than 4 are considered without internal excitation energy; 
larger fragments are allowed to carry an excitation
energy ($\epsilon$) upper limited by the binding energy according
to the following level density formula:

\begin{equation}
\rho(\epsilon)=\frac{\sqrt{\pi}}{12 a^{1/4}\epsilon^{5/4}}
\exp(2 \sqrt{a \epsilon}) \exp(-\epsilon/\tau),
\end{equation}
with $a=0.114 A+0.098 A^{2/3}$ MeV$^{-1}$ \cite{Iljinov} and $\tau$=9 MeV.
The factor $\exp(-\epsilon/\tau)$ takes into account the decrease of the excited
level lifetime with the increasing excitation energy.

The excited primary fragments obtained in the break-up stage are allowed to decay by 
sequential particle emission (second stage of the model) as in Ref. \cite{Noi_Indra}. 
The range of the evaporated particles is considered up to $A=16$.
As already stated in literature \cite{Souza,Tan} by modifying isotopic yields, 
secondary particle production
mechanisms may alter isoscaling parameters. 

Despite the fact that the employed parameterizations are expected to affect the specific values of 
$\alpha$ and $\beta$ they do not affect the general behavior of isoscaling parameters as a function of
observables characterizing the sources' state such as the conclusions of the study are general and valid.

\subsection{General microcanonical results}

In order to verify to what extend isoscaling holds within a microcanonical
framework and, if this is the case,
to make a systematic study of all source parameters which might have
an influence on the isoscaling parameters we considered pairs of equilibrated
sources with masses ranging from $A=40$ to $A=200$, excitation energies ranging
from 2 to 15 MeV/nucleon, freeze-out volumes ranging from $3V_0$ to $10V_0$ 
in both primary and asymptotic stages of the reaction. 

For all considered cases we obtained isoscaling in the sense that
$\log R_{21}(Z,N)$ shows an almost perfect linear dependence 
as a function of $N$ and $Z$. 

As general observations, one may say that the quality of the isotopic
scaling is better in the break-up stage than in the asymptotic stage 
of the decay and the maximum size of the emitted cluster for which
isoscaling still holds decreases with increasing excitation energy. 
We think that the more modest quality of the isoscaling in the asymptotic 
stage is an artifact
of the simplified manner in which the secondary decays have been implemented.
We adopt the widely used convention to denote with the index "2" the more 
neutron rich system and with the index "1" the more neutron poor system. 
In this situation the value of $\alpha$ is
always positive because more neutron rich clusters will be produced by the
neutron richer source and the value of $\beta$ is always negative.
To avoid possible effects of different magnitudes
of Coulomb interaction on isotopic distributions, the considered pairs of sources 
have the same proton number and different mass numbers. 

\begin{figure}

\resizebox{0.8\textwidth}{!}{%
  \includegraphics{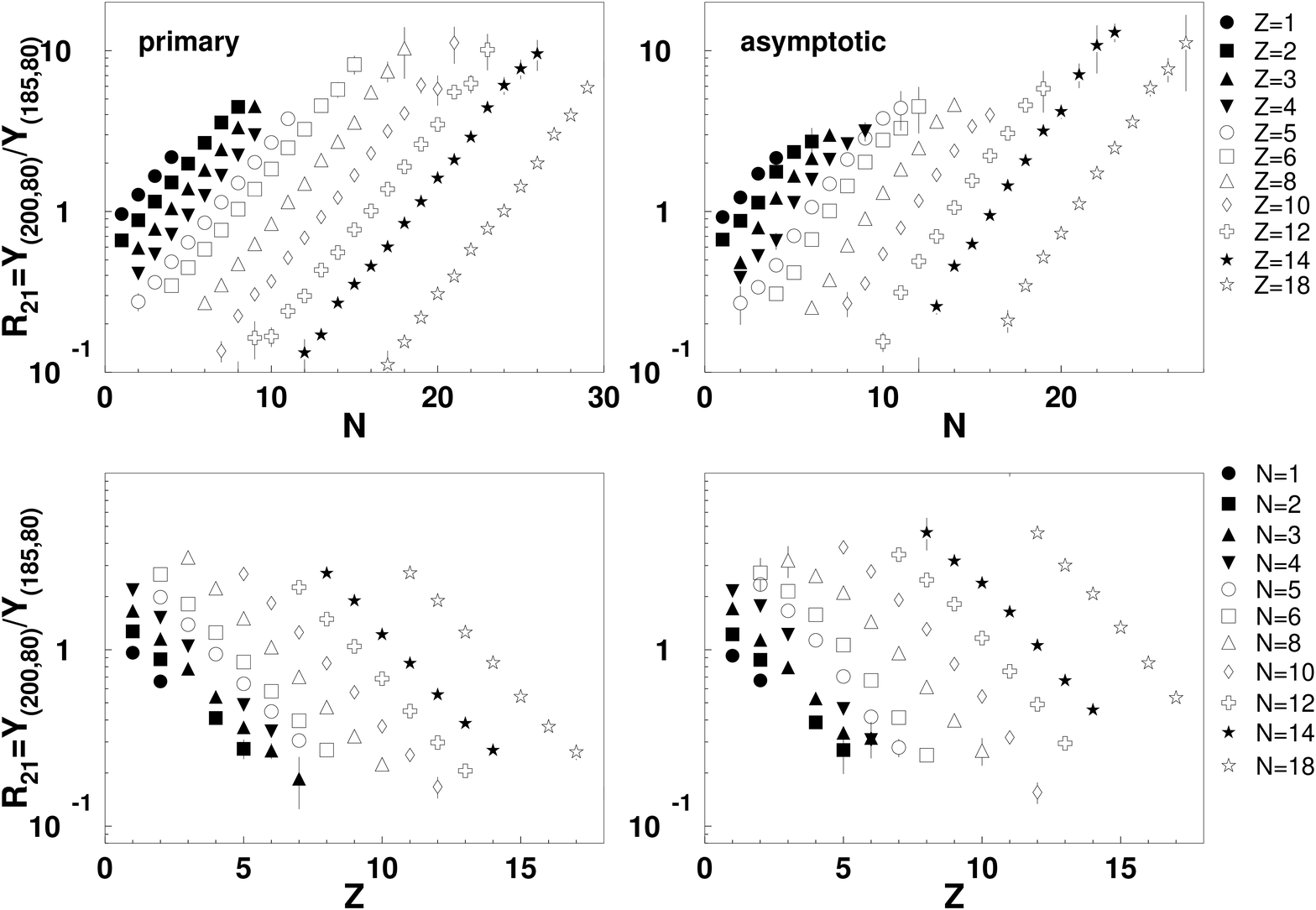}}

\caption{Isotopic yield ratios are plotted in logarithmic scale
as a function of $N$ (upper panels) and
$Z$ (lower panels) for two equilibrated sources (200, 80) and (185, 80) with the
excitation energy $E_{ex}$=5 MeV/nucleon and the freeze-out volume $V=6V_0$ in both
primary (left panels) and asymptotic stages of the decay (right panels).
The considered sequence of $N(Z)$ is 1,2,3,4,5,6,8,10,12,14,18.}
\label{fig:z=80}
\end{figure}

To illustrate how isoscaling looks like in the case of this microcanonical
model, in Fig. 1 we present the isotopic yield ratios as a function of 
$N$ and $Z$ for the pair of equilibrated sources (200, 80) and (185, 80) at
5 MeV/nucleon excitation energy and a freeze-out volume 6 times larger
than the nuclear volume at normal density. 
As results from the figure, isoscaling holds in the strict sense of Eq. 
\ref{eq.isoscaling}, namely linear behavior of $\log R_{21}$ as a function of 
$N$ and $Z$ and constant value of $C$. 
Special attention should be drawn to the fact that in Fig. \ref{fig:z=80}
distances from the points corresponding to different values of $Z(N)$ is not
equal because the considered sequence of $Z(N)$ does not contain exclusively
consecutive numbers (see legend).
 
It is interesting to notice that
a linear behavior of $\log R_{21}$ is obtained also for emitted clusters much
heavier than those for which the experimental analysis has been possible. 
As a technical detail, we stress that despite
for a given element the linear behavior of $\log R_{21}(N)$ holds for almost all
obtained isotopes, for the fitting procedure only the most stable fragments have
been selected. 

\subsection{Dependence of $\alpha$ and $\beta$ on excitation energy}

Since, as known from the early days of multifragmentation, the excitation energy of the 
nuclear source induces strong modifications on all fragment multiplicities, it is expected to 
affect also isotopic yield ratios. 

\begin{figure}
\resizebox{0.5\textwidth}{!}{%
  \includegraphics{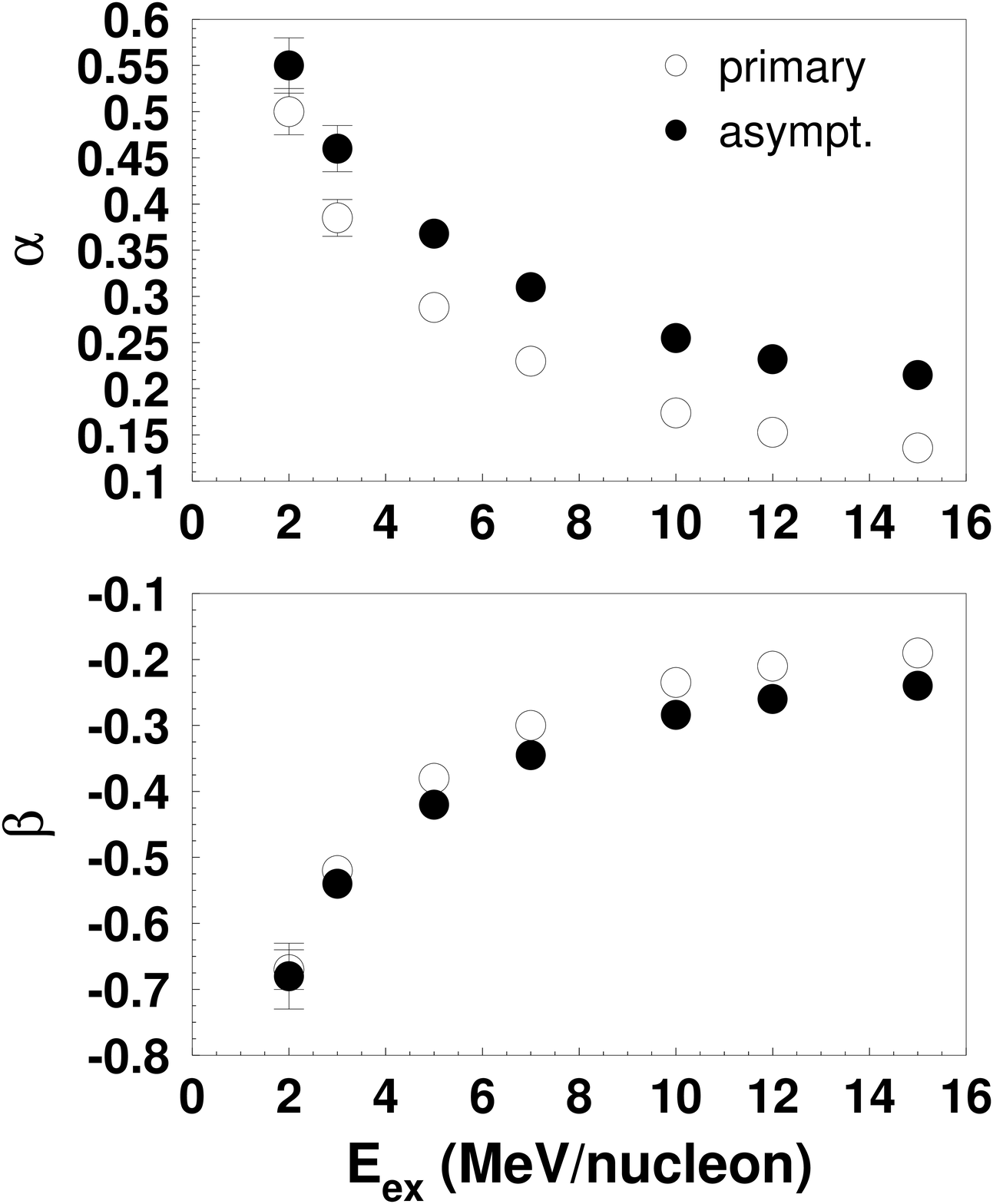}
}
\caption{Slope parameters $\alpha$ and $\beta$ as a function of
excitation energy for primary (open symbols) and asymptotic (full
symbols) stages of the decay. The considered
equilibrated sources are (200, 80) and (185, 80). The freeze-out volume is 
$V=6V_0$.}
\label{fig:ab_ex}
\end{figure}

In Fig. \ref{fig:ab_ex} are represented $\alpha$ and $\beta$ for the equilibrated sources 
(200, 80) and (185, 80), with $V=6V_0$ as a function of the excitation energy.
It must be specified that, if not mentioned in a different manner,
along this paper $\alpha$ and $\beta$ are
calculated such as to obtain the best fit for fragments with $2<Z<9$ and 
respectively $2<N<9$. This choice is made
in order to realize an analysis as much as possible
similar to the way in which the slope parameters have been obtained from
experimental data and to have the same amount of considered data for all
excitation energies and all considered source sizes.

One can see that both parameters have a monotonic dependence on the 
excitation energy, their absolute value decreasing with the temperature.
Thus, increasing the excitation energy from 2 MeV/nucleon to 15 MeV/nucleon 
$\alpha$ diminishes by a factor of 2.5 and 
$|\beta|$ diminishes by a factor of 3.5
This effect can be understood having in mind that an increase in 
excitation
energy will result in a more uniform population of the isotopes corresponding to a
given element thus washing out the effect of the isotopic 
composition of the source. 
In terms of Eqs. (\ref{eq.alpha.g}) and (\ref{eq.beta.g}), 
an increasing excitation energy will lead to a shift of the $Y(x,y)|_y$ 
distributions
toward smaller values of $x_{{\sf med}}$ such as
$(x_{{\sf med}2}-x_{{\sf med}1})$ slightly diminishes together with 
an enhancement of the distributions' variance $\sigma$. Since both numerator and denominator
from Eqs. (\ref{eq.alpha.g}) and (\ref{eq.beta.g}) act in the same sense, the decrease of  
$\alpha$ and $|\beta|$ with the excitation energy of the sources is obvious.

These results are in qualitative agreement with those of Ref. \cite{Botvina,shetty}.

\subsection{Effect of secondary decays on $\alpha$ and $\beta$ slope parameters}

An important problem that manifests whenever we need to infer the break-up
information from the available experimental multifragmentation data is 
the reconstruction of primary observables affected by the secondary particle
emission. This problem may be even more important when one operates with very 
sensitive quantities like the slope parameters $\alpha$ and $\beta$.
It was stated in the literature that when the two considered sources
have almost identical parameters excepting the isospin, the effects of 
sequential evaporation cancel one another
leading to small modifications of the break-up stage results
\cite{Tsang_RC2001}. On the other hand, estimations performed with accurate 
models like SMM
\cite{SMM} indicate that modifications of the slope parameters due to
evaporation may manifest with different magnitudes depending on how
sequential evaporation is implemented \cite{Botvina,Tsang_stat_mod}.

To check the evaporation effect on isoscaling within the present model in
Fig. \ref{fig:ab_ex} are plotted
$\alpha$ and $\beta$ in the break-up and asymptotic stages 
of the decay
for the equilibrated sources (200, 80) and (185, 80), with $V=6V_0$ as
a function of the excitation energy. 
As displayed by the figure, over the entire energy range the absolute values of 
$\alpha$ and $\beta$ are larger in the asymptotic stage of the decay 
in comparison to the break-up stage.
 
To get an insight on what happens during the sequential particle emission 
from primary excited fragments is necessary to focus on the evaporation process.
Given the fact that for all excitation energies neutrons are emitted with the most
important probability, the final cold fragments are more symmetric than their 
break-up ancestors. This means that the isotopic distribution of a given $Z$ shifts
toward smaller values of $N_{med}$. Since the amplitude of this shift is larger for
the neutron richer source, it means that 
the value $\left(N_{{\sf med}2}(Z)-N_{{\sf med}1}(Z)\right)$ in the asymptotic stage
is smaller than the one corresponding to the break-up stage. This apparent
trend of isoscaling parameters to decrease after particle evaporation is annihilated by
a strong narrowing of isotopic distributions (the more asymmetric nuclei have small
survival probabilities after secondary decays) such as in the end the values of
$\alpha$ and $|\beta|$ increase with respect to their break-up values. This result is
counterintuitive because at a first glance one could expect secondary decays to wash out
the isotopic differences of the sources.

From Fig. \ref{fig:ab_ex} is obvious that secondary decays affect more 
$\alpha$ than $\beta$. This result can be understood taking into account that 
the decrease of the width of $Y(Z,N)|_Z$ distributions caused by
neutron evaporation is more important than the narrowing of $Y(Z,N)|_N$ 
distributions caused by less important charged particle emission. 

\begin{figure}
\resizebox{0.5\textwidth}{!}{%
  \includegraphics{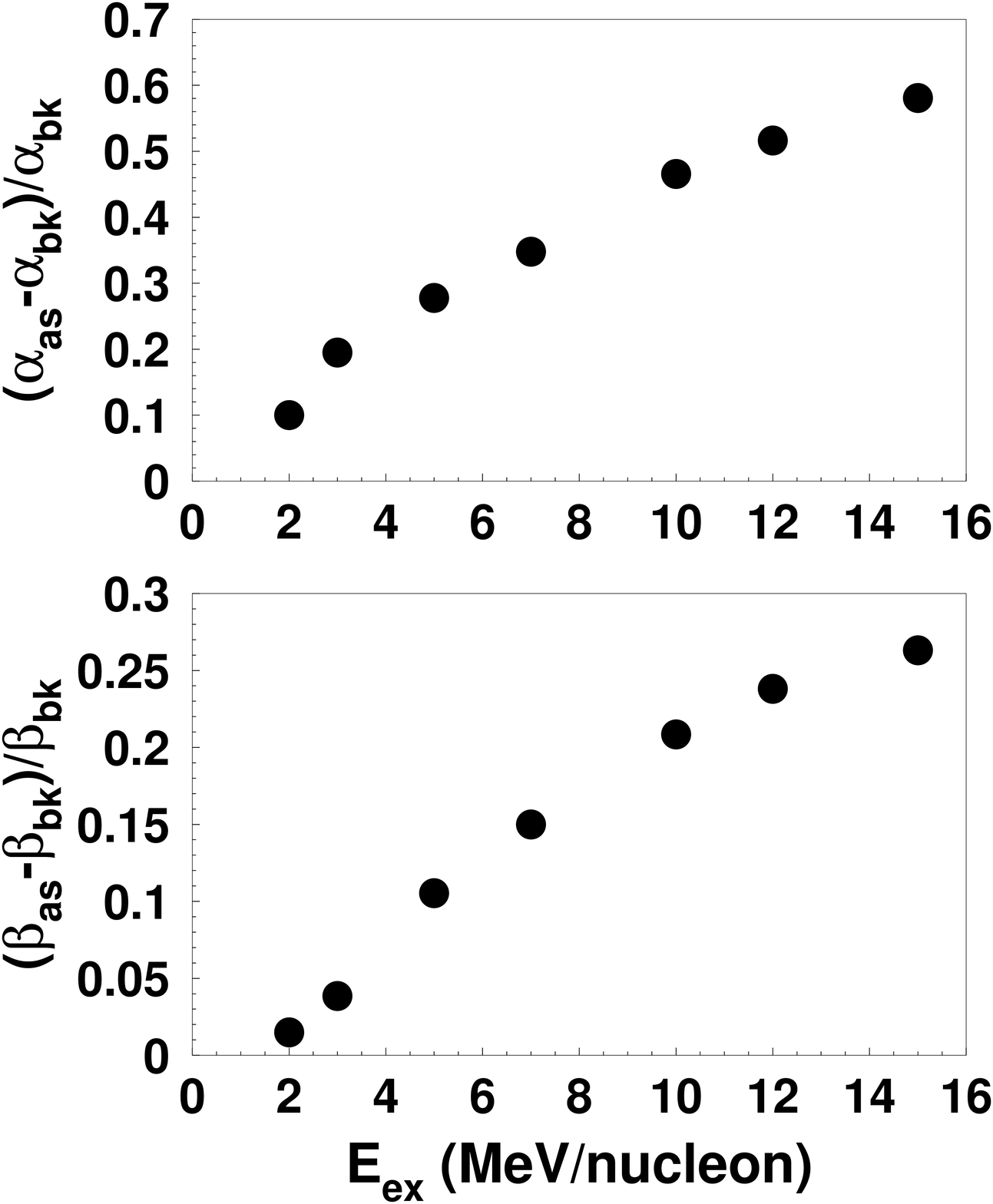}
}
\caption{Effect of secondary decays on the slope parameters $\alpha$ and $\beta$ 
function of excitation energy.
The considered equilibrated sources are (200, 80) and (185, 80) with $V=3V_0$.}
\label{fig:rap_ab_ex}
\end{figure}

One can see from Fig. \ref{fig:ab_ex} that
the difference between the asymptotic and break-up values increases up to 
5 MeV/ nucleon
excitation energy and then stays roughly constant, but taking into account 
that the increase
in excitation energy results in smaller slope parameters, this means that the 
relative 
influence of the secondary decays on $\alpha$ and $\beta$ grows with energy.
To illustrate the relative modifications brought by the secondary decays
to the slope parameters in Fig. \ref{fig:rap_ab_ex}
is plotted the deviation between asymptotic and
primary values relative to the break-up results versus excitation energy. The
increasing rate is roughly constant for whole considered domain.

\subsection{Slope dependence on the freeze-out volume}

To identify a possible freeze-out volume effect on the slope parameters,
isoscaling was analyzed for the previously mentioned pair of sources 
((200, 80) and (185, 80)) at excitation energies increasing from 2 to 15
MeV/nucleon at $V=3V_0$, $V=6V_0$, $V=10V_0$ in both primary and asymptotic
stages of the decay. 
Despite the fact that all isotopic distributions present a strong dependence on the 
volume of the excited source,
for the considered cases no significant freeze-out volume dependence
of $\alpha$ and $\beta$ was identified. 

The impossibility to get an analytic expression of isotopic yields within a 
microcanonical model makes very hard to estimate the effect of freeze-out volume 
(or any other observable characterizing the state of the source) on isotopic
ratios. However,
if true, the independence of the slope parameters on the freeze-out volume is of
particular importance for experiments aiming to extract the symmetry term of the binding energy from
isotopic yields. Indeed, no matter is the method for selecting collisions, is hard to believe that
the ensemble of equilibrated nuclear systems formed in heavy ion reactions correspond to one value
of the freeze-out volume, but rather to a distribution. Since there is no method to select fragmentation 
events according to their volume, volume dependent $\alpha$ and $\beta$ would make impossible
the extraction of the asymmetry term.

Ref. \cite{Tsang_stat_mod} makes use of a canonical version of SMM \cite{SMM_gc} and reaches
the conclusion that for all considered temperatures ($T$=4, 5, 6 MeV) $\alpha$ increases monotonically with the
freeze-out density: by increasing $\rho/\rho_0$ from 0.1 to 1 $\alpha$ increases by a factor of two.
The opposite conclusions of Ref. \cite{Tsang_stat_mod} and the present study are produced by the 
in principle
different basic hypothesis of each model. Taking into account the implications of the volume dependence
of isoscaling parameters, a definite answer to this problem would be of much interest.


\subsection{Slope dependence on the considered sources}

All the above discussion suggests a strong dependence of the fragment production on the
parameters characterizing the equilibrated state of the sources. A natural question to rise
at this point is how sensitive are the slope parameters on the considered sources once one
keeps fixed the atomic number, the freeze-out volume and the excitation energy.

\begin{figure}
\resizebox{0.55\textwidth}{!}{%
  \includegraphics{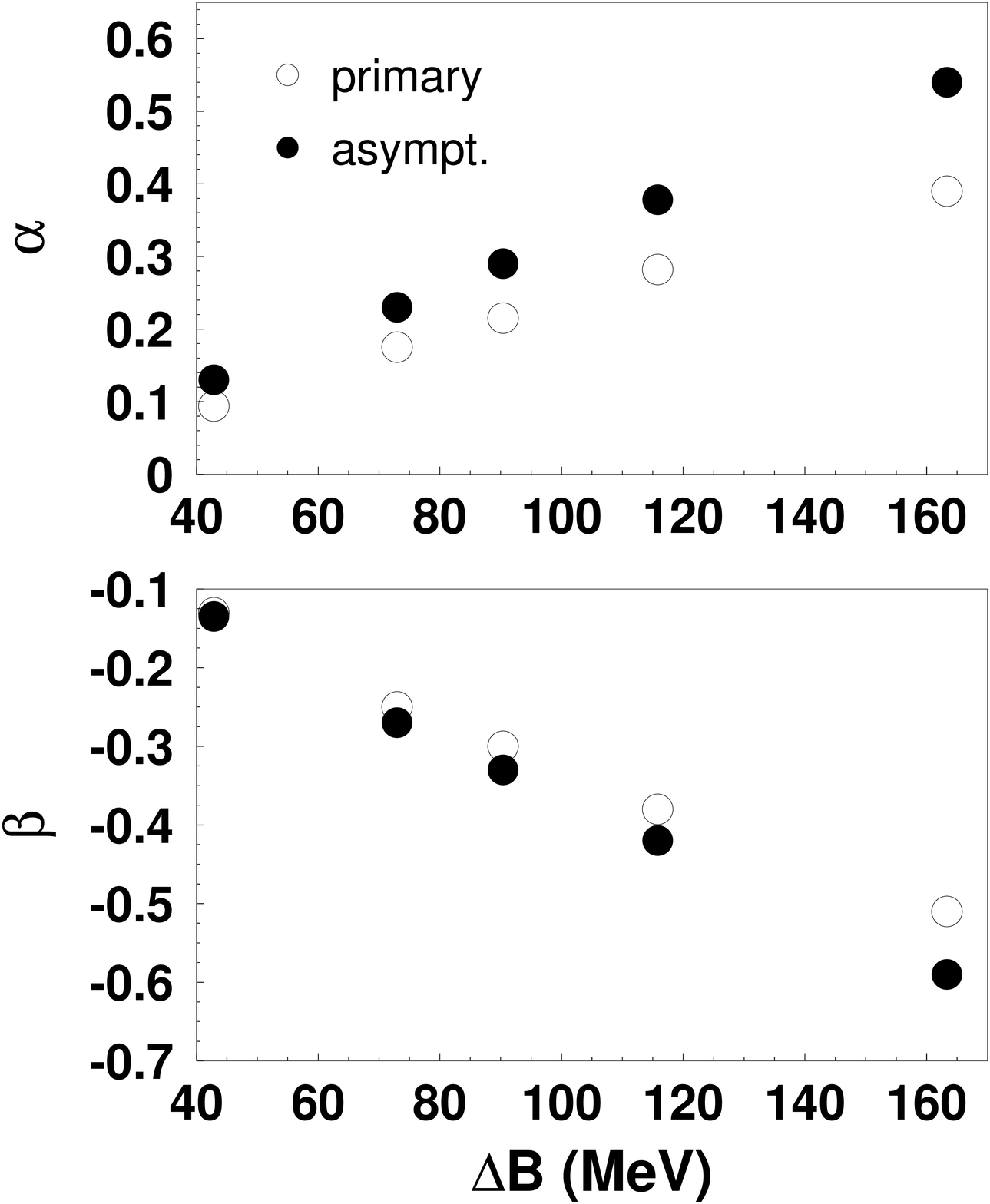}
}
\caption{Slope parameters $\alpha$ and $\beta$ as a function of the difference
between the binding energies of the equilibrated sources.  
Open symbols correspond to break-up isotopic yields and full symbols correspond
to isotopic yields after sequential particle emission.
The considered equilibrated sources are characterized by $Z=80$, 
$E_{ex}=5$ MeV/nucleon and $V=3V_0$.}
\label{fig:ab_b}
\end{figure}

To answer this question $\alpha$ and $\beta$ have been
calculated for different pairs of sources
(chosen from the set (180,80), (185,80), (190,80) and (200,80)) at 
$E_{ex}=5$ MeV/nucleon and $V=3V_0$.
The result we obtained is that the slope parameters are not dictated by the difference of 
isospin asymmetry between the two sources, but by the difference between the binding
energy of the sources. To illustrate this result we plot in Fig. \ref{fig:ab_b}  
the monotonic dependence of $\alpha$ and $\beta$ as a function of the difference
between the binding energies of the equilibrated sources.

This result is in qualitative agreement with some experimental data. 
Ref. \cite{Liu} reports a modification of the absolute values of $\alpha$ and $\beta$ 
by a factor of two when calculated using ($^{112}$Sn+$^{124}$Sn/ $^{112}$Sn+$^{112}$Sn) 
nuclear reactions
instead of ($^{124}$Sn+$^{124}$Sn/ $^{112}$Sn+$^{112}$Sn) at 50 MeV/nucleon. Similar
results were obtained recently studying isoscaling in 
($^{58}$Fe+$^{58}$Ni/ $^{58}$Ni+$^{58}$Ni)
and ($^{58}$Fe+$^{58}$Fe/ $^{58}$Ni+$^{58}$Ni) 
reactions at 30, 40 and 47 MeV/nucleon \cite{shetty}.


Contrary to the cases of a grandcanonical model or a sequential particle emission model
where one can easily express isotopic multiplicities as a function of the binding energy of the 
parent nucleus, within a microcanonical model the effect of binding energy can be estimated only
in an intuitive way.
Thus, more similar are the two considered equilibrated sources from the point of view of their binding energies,
closer are the values of the total available energy of each system for a given value of the excitation energy.
This will lead
to similar populations of the configuration space of each system. In this case, all observables (including isotopic
yields) are expected to have close values. Close values of isotopic yields will obviously produce small
values of $\alpha$ and $\beta$.

\subsection{Slope dependence on the source size}

The only parameter characterizing the equilibrated source which was kept fixed thorough 
the above discussion is the source's size. To check the size effect of isoscaling parameters 
we choose for fragmenting sources two pairs of smaller nuclei, 
namely ((112, 50), (124, 50)) and ((112, 50), (119, 50)) with $E_{ex}=5$ MeV/nucleon 
and $V=3V_0$. 
The values of the slope parameters obtained
in this case (((112, 50), (124, 50)):
$\alpha_{bk}$=0.37, $\alpha_{as}$=0.50, $\beta_{bk}$=-0.49 and $\beta_{as}$=-0.55;
((112, 50), (119, 50)):
$\alpha_{bk}$=0.23, $\alpha_{as}$=0.32, $\beta_{bk}$=-0.30 and $\beta_{as}$=-0.34)
are very different from the values obtained for the $Z=80$ sources and
suggest a strong dependence of the isoscaling parameters on source size.

This result is in contradiction with Ref. \cite{Tsang_stat_mod} where the canonical version of 
SMM \cite{SMM_gc} predicted $\alpha$ independent of the size of the sources. The discrepancy between
the cited canonical results and the present microcanonical results focuses the attention on the 
importance of using appropriate models for describing finite systems.

\begin{figure}
\resizebox{0.8\textwidth}{!}{%
  \includegraphics{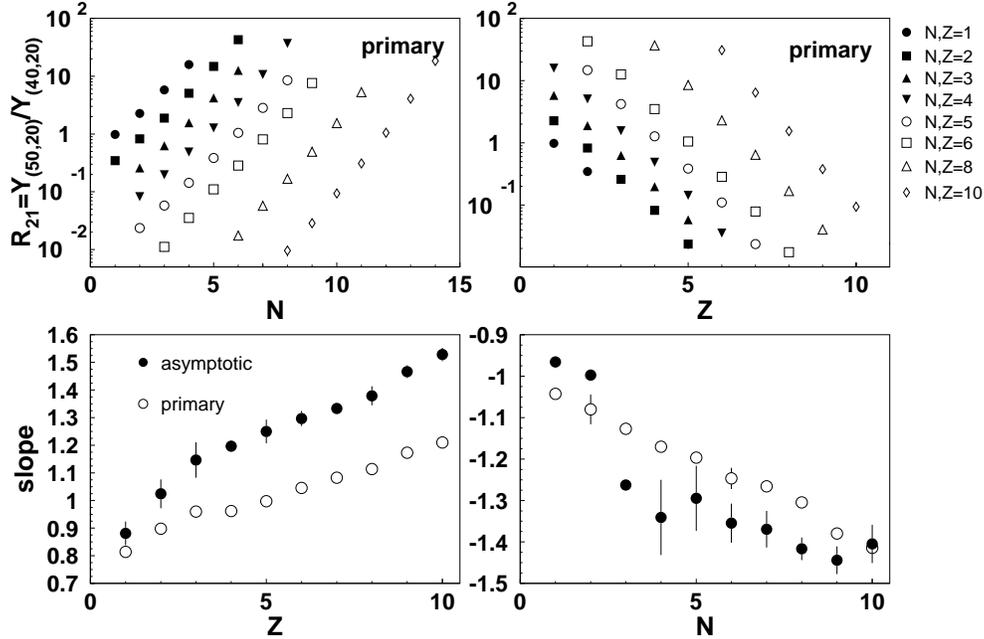}
}
\caption{Upper panels: Isotopic yield ratios for the primary decay of (50, 20) 
and (40, 20) at $E_{ex}=5$ MeV/nucleon and $V=6V_0$
as a function of $N$ (left panel) and $Z$ (right panel). 
Lower panels: Dependence of the slope parameters $\alpha$ and $\beta$ of the
size of the considered emitted cluster. Here $\alpha$ ($\beta$)
represents the slope of $\log R_{21}(Z,N)$ versus $N$ ($Z$) for each
considered value of $Z$ ($N$).
The considered sequence of $N(Z)$ is 1,2,3,4,5,6,8,10.}
\label{fig:z=20}
\end{figure}

\subsection{Slope dependence on the emitted fragment size}

Taking into account that isoscaling has been
argued on grandcanonical basis and above presented microcanonical results suggest important
deviations from grandcanonical expectations, it would be interesting to check
whether isoscaling is valid also for smaller sources. 

The results obtained for ((40, 20) and (50, 20)) at $E_{ex}=5$ MeV/nucleon and 
$V=6V_0$ proved a linear behavior
of the yield ratios with the isotopic composition of the emitted fragment together with a
monotonic increase of the absolute value of the slope parameters with the size of the emitted
fragment as displayed in Fig. \ref{fig:z=20}
where we plot the results obtained for fragments with $1 \leq Z \leq 10$ and
$1 \leq N \leq 10$. 
This means that for small sources we
obtain isoscaling in the sense of Eqs. (\ref{eq.alpha.g}) and (\ref{eq.beta.g})
without the constancy of the slope parameters with the size of the cluster.

\begin{figure}
\resizebox{0.55\textwidth}{!}{%
  \includegraphics{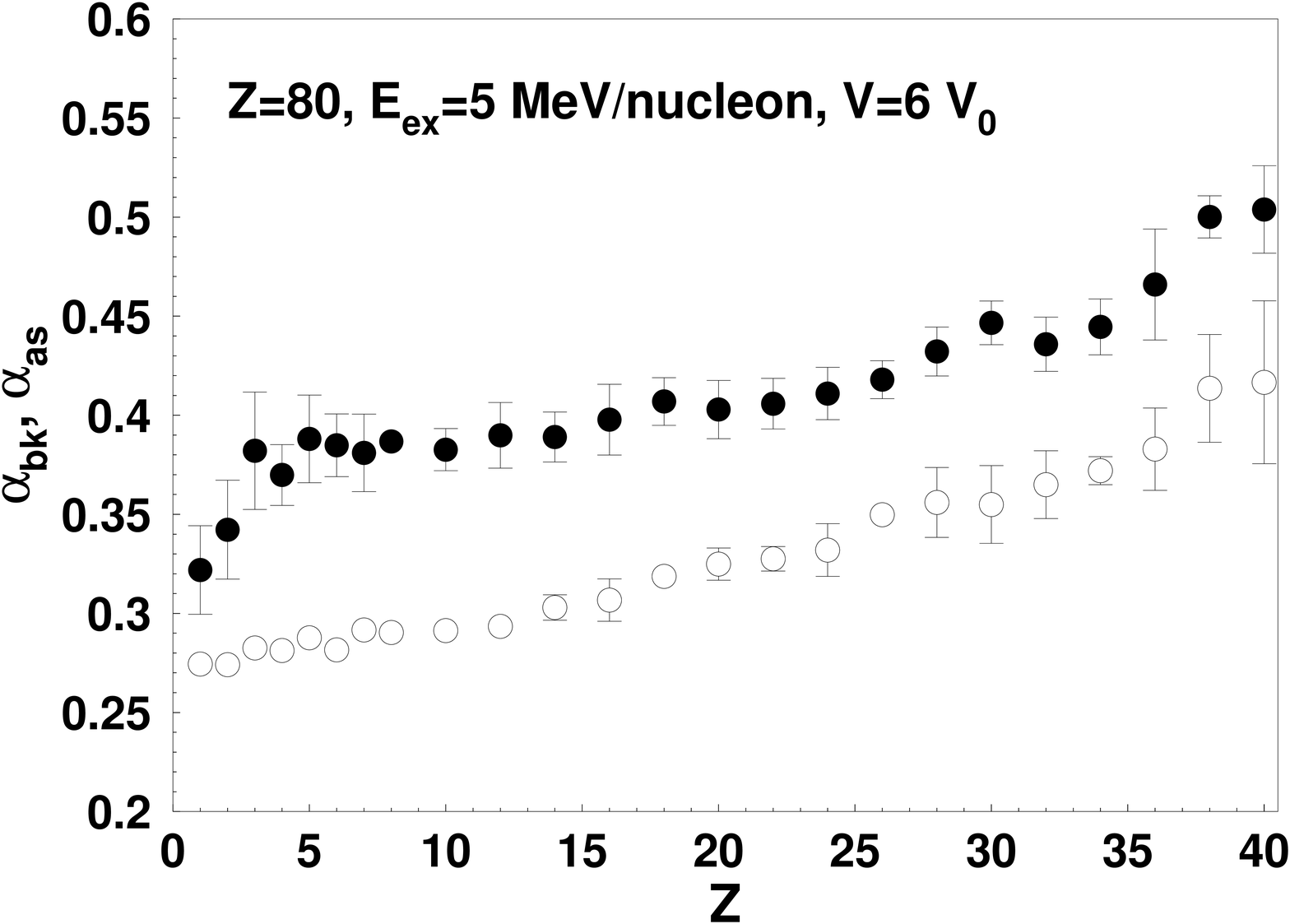}
}
\caption{Variation of the slope parameter $\alpha$ with $Z$ 
for the
fragmenting systems (185,80) and (200,80) at $V=6V_0$ and 
$E_{ex}=5$ MeV/nucleon.
Open symbols correspond to the primary
stage, close symbols correspond to the asymptotic stage of the decay.}
\label{fig:fit_z=80}
\end{figure}

This result makes us go back to large fragmenting sources in order to evidence the
dependence of $\alpha$ and $\beta$ on the size of the emitted cluster. From our analysis
results that by considering large emitted fragments isoscaling is obtained with slope 
parameters
dependent on the size of the cluster. To illustrate this result, in Fig. \ref{fig:fit_z=80}
we plot $\alpha(Z)$ for ((200, 80) and (185, 80)) at $V=6V_0$ and $E_{ex}=5$ MeV/nucleon.
One can see that while for $Z<10$ $\alpha$ is constant, for larger emitted fragments 
$\alpha$ increases monotonically with $Z$. 
It is interesting to mention that isoscaling slope parameters dependent on the emitted cluster size 
have been evidenced also in the framework of a dynamical stochastic mean field model \cite{Liu}.

Because present detectors are not able to isotopically separate fragments with charge larger than 8,
the only case when one may hope to evidence this effect corresponds to light fragments 
emitted by light sources.
 
The fragment dependence of the slope parameters can be interpreted as a
finite size effect which, as expected, enhances with the diminish of the source
and with the increase of the emitted fragment.

\section{Conclusions}
\label{sec.concl}

To summarize, the present paper aims to investigate isoscaling in nuclear multifragmentation.
In order to offer a complete and as accurate as possible image on the subject, the study was done 
using a microcanonical multifragmentation model and covered
a large range of source sizes ($A$=40-200), excitation energies ($E_{ex}$= 2 - 15 MeV/nucleon)
and freeze-out volumes ($V$= 3$V_0$ - 10$V_0$). Primary and asymptotic
stages of the decay have been considered.

The results indicate that despite in all considered cases the logarithm of the isotopic yields corresponding
to the two different sources have a linear behavior as a function of $Z$ and $N$, the slope parameters are strongly
dependent on most observables characterizing the state of the sources. The only observable not affecting 
$\alpha$ and $\beta$ is the freeze-out volume. This result could be of particular importance for experiments
aiming to extract the asymmetry term of the binding energy using multifragmentation data, since most probably
equilibrated excited systems formed in heavy ion collisions have a distribution of volumes and the event separation
as a function of volume is not by far possible.
The isoscaling slope parameters are affected also by sequential particle emission.

Deviations from strict isoscaling have been observed for heavy clusters originating from 
large multifragmenting systems
and light clusters emitted by light systems and can be interpreted as finite size effects. 
Such deviations from
grandcanonical predictions are expected to manifest especially for small systems and stress the importance of
using approximations suitable to the physical situation.

Since the microcanonical ensemble is by principle the most appropriate statistical tool to describe
the disintegration of small isolated systems and microcanonical multifragmentation models in general
proved to be able to describe with accuracy experimental data, 
the most reliable description of isoscaling is expected to be obtained with this kind of models.
The accuracy of isoscaling parameters is essential for obtaining correct information on fundamental quantities like
the asymmetry term of the binding energy.

\end{document}